\documentclass{aa}

\usepackage{booktabs}
\usepackage{graphicx}
\usepackage{enumitem}
\graphicspath{{figs/}}
\usepackage{hyperref}
\usepackage[all]{hypcap}
\usepackage[capitalise]{cleveref}
\usepackage[separate-uncertainty]{siunitx}
\usepackage{placeins}
\usepackage{txfonts}
\usepackage{xcolor}
\usepackage{ulem}
\usepackage{ifthen}

\begin{document}

   \title{A study of the large-scale formation in the environment of A3266}
   \subtitle{Infalling groups, filaments, and a premerger cold front}

   \author{J. Dietl\inst{1,2,3}
          \and A. Veronica\inst{1,2}
          \and T. H. Reiprich\inst{1,2}
          \and F. Pacaud\inst{1,2}
          \and Y. Zhao\inst{1,2}
          \and J. S. Sanders\inst{4}
          \and B. Seidel\inst{5}
          \and M. C. H. Yeung\inst{4}
          \and K. Dolag\inst{5,6}
          \and E. Gatuzz\inst{4}
          }

   \institute{Argelander-Institut für Astronomie (AIfA), Universität Bonn, Auf dem H\"ugel 71, 53121 Bonn, Germany \\ e-mail: \texttt{jdietl@uni-bonn.de}
   \and
   Cluster of Excellence DYNAVERSE
   \and
   Max-Planck-Institut für Radioastronomie (MPIfR), Auf dem Hügel 69, 53121 Bonn, Germany
   \and
   Max-Planck-Institut für extraterrestrische Physik (MPE), Gießenbachstraße 1, 85748 Garching, Germany
   \and
   Universitäts-Sternwarte, Fakultät für Physik, Ludwig-Maximilians-Universität München, Scheinerstr. 1, 81679 München, Germany
   \and
   Max-Planck-Institut für Astrophysik, Karl-Schwarzschild-Straße 1, 85741 Garching, Germany}

   \date{Received 8 May 2026 / Accepted 29 June 2026}

  \abstract
   {Abell~3266 (A3266) is a dynamically active galaxy cluster embedded in a dense environment of galaxy groups and clusters at similar redshift. Data from the \textit{Spektrum Roentgen Gamma (SRG)/eROSITA} all-sky survey enable us to study the large-scale emission with a low X-ray surface brightness in the outskirts of galaxy clusters.}
   {We investigate the unexplored outskirts of A3266 out to $3R_{100}$, characterize its closest neighboring galaxy group, and search for connecting filaments using X-ray emission and galaxy number density.}
   {We performed a detailed X-ray image, surface brightness, and spectral analysis in selected regions and sectors. The X-ray analysis was complemented by the distribution of member galaxies from the \textit{NASA/IPAC Extragalactic Database NED} and by a comparison with the cosmological simulation \textit{Simulating the LOcal Web (SLOW)}.}
   {An X-ray emission filament, connecting A3266 to its closest neighboring group in the northwest and extending over a 3D length of $L_{R_{200}\text{–}R_{200}} = {1.1}_{-0.1}^{+0.5}\,\mathrm{Mpc}$, is detected with a significance of $3.6\,\sigma$. The group exhibits cool-core properties and is embedded within the filament. The filament temperature is $T=\num{1.2}_{-0.2}^{+0.3}\,\mathrm{keV}$, the metallicity is  $Z = \num{0.07}_{-0.05}^{+0.09}\,Z_{\odot}$, and for a simple geometry, we derived its electron number density as $n_{\rm e} = {8}_{-2}^{+1}\times10^{-5}\,\mathrm{cm}^{-3}$ .}
  {Our findings reveal a coherent network of galaxy groups in the outskirts of A3266, tracing the ongoing assembly of the cluster along large-scale structure. The detected filament is hotter and denser than expected for the pristine warm-hot intergalactic medium, consistent with gas processed in the cluster environment and further affected by the infall of the northwest group.  The comparison with SLOW shows that the observed group distribution and filamentary connections are qualitatively consistent with expectations for an actively accreting cluster embedded in a large-scale structure.}


   \maketitle
%

\section{Introduction}
\label{sec:introduction}
Abell~3266 (A3266), located in the southwestern sky, is a nearby massive galaxy cluster at a redshift of $z=\num{0.0596}$ \citep{Dehghan_2017}. Within $R_{500}$, A3266 has been extensively studied at multiple wavelengths. For instance, \citet{Finoguenov_2006} used observations from the \textit{XMM-Newton} telescope to confirm a merger state in the northeast–southwest direction and found an elongated and asymmetric intracluster medium (ICM), as revealed by entropy, pressure, and temperature maps. The optical analysis by \citet{Dehghan_2017} identified six additional substructures surrounding the core, which itself is split into a northeast and a southwest component, consistent with the merger direction inferred from X-ray observations. \textit{eROSITA} observations presented by \citet{Sanders_2022} identified X-ray counterparts of several of these substructures and revealed a highly asymmetric X-ray surface brightness distribution within $R_{200}$, including two weak shocks. Temperature and metallicity maps further confirmed the structures reported by \citet{Finoguenov_2006}. All three studies concluded that A3266 is not a simple binary merger, but a highly complex and dynamically active system with an unusual amount of internal substructure.

The line-of-sight velocity structure of the ICM was studied by \citet{Gatuzz_2024}. They found lower gas velocities than in the merging substructures measured by \citet{Dehghan_2017}, but still emphasized the significant dynamical interactions that shaped the cluster velocity field. They also reported a correlation between velocity and metallicity structure, with blueshifted emission in the low-metallicity northwestern region and redshifted emission in the high-metallicity southeastern region relative to the cluster mean redshift. \citet{Gatuzz_2025} presented radial abundance profiles within $R_{500}$ and reported discontinuities following temperature discontinuities, likely caused by the subgroups.
Radio observations revealed additional complexity, including a wrong-way radio relic, two ultra-steep-spectrum fossil plasma sources, an extended central diffuse ridge, and a large-scale radio halo (e.g., \citealt{Rudnick_2021}, \citealt{Duchesne_2022}, \citealt{Riseley_2022}; see Fig.~1 in the latter for an overview of the radio sources).

\begin{figure}[htbp]
    \centering
    \includegraphics[width=\hsize]{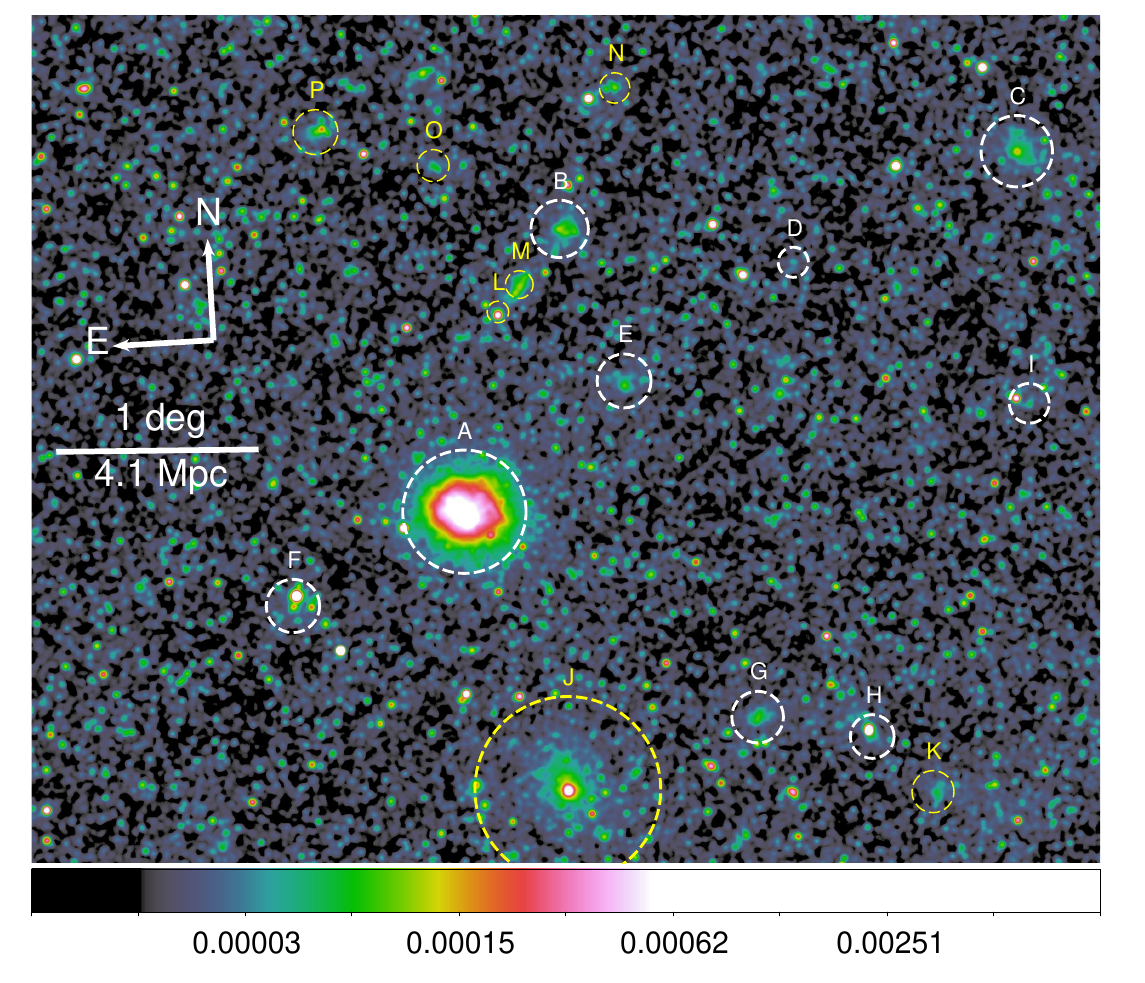}
    \caption{eROSITA image of the large-scale field around A3266 in the soft X-ray band. The data reduction steps to produce this image are explained in \cref{sec:data_reduction}. The color bar is in units of counts per second (as in all following eROSITA images). The circles show known galaxy groups and clusters with their $R_{500}$ from the eROSITA DR1 catalog \citep{Bulbul_2024}, color-coded dependin on when they are within a $\SI{50}{Mpc}$ line-of-sight difference to A3266 (white) or not (yellow), corresponding to a redshift range of $z = 0.0476$--$0.0716$ (if interpreted purely cosmologically). The names and redshifts of all marked sources can be found in \cref{tab:group_props}. Source A is A3266 itself, and source C is the galaxy cluster A3255. The line-up of sources F, A, E, D, and C points toward the direction of the Horologium-Reticulum supercluster in the NW. To be consistent with all $R_{500}$ radii in this image, we show the eROSITA catalog value for A3266, although we use the $R_{500}$ radius of \citet{Ettori_2019} in the rest of the paper (\cref{tab:cluster_props}), which is $\SI{12}{\percent}$ larger.}
    \label{Fig:large-scale_overview}
\end{figure}

Our large-scale \textit{eROSITA} overview image (\cref{Fig:large-scale_overview}) shows A3266 (source A) and its surrounding galaxy groups and clusters, color-coded depending on when they are within a 50 Mpc line-of-sight difference to A3266 (white) or not (yellow). The second most massive cluster in the field is A3255 (source C). A3266 is located at the southwestern edge of the Horologium–Reticulum supercluster (HRSC; e.g., \citealt{Lucey_1983}, or Fig.~9 in \citealt{Sifon_2025} for a visual overview). The sources F, A, E, D, and C align in the direction of A3164 ($\approx\SI{3}{\deg}$ outside of the image section in NW direction), which lies at the same redshift and is one of the core clusters of the HRSC. Furthermore, sources A, B, and C are identified as members of a common supercluster in the eROSITA supercluster catalog \citep{Liu_2024}.

These observations together indicate that A3266 has experienced a complex merger history and is still undergoing active merging and mass accretion from its surrounding large-scale environment. This makes A3266 an ideal target for studying the interface between cluster outskirts and the cosmic web. We present the first study of the outskirts of A3266 beyond $R_{200}$, focusing on its connection to the surrounding galaxy groups.

\begin{table}[htbp]
\centering
\caption{General properties of A3266.}
\label{tab:cluster_props}
\begin{tabular}{lc}
\toprule
RA, Dec [FK5; J2000] & $\SI{67.850}{\degree}$, $\SI{-61.444}{\degree}$\ \textsuperscript{a}\\
redshift $z$ & $\num{0.0596(2)}$\ \textsuperscript{b} \\
velocity dispersion $v_{\rm disp}$ & $\SI{1337(67)}{km/s}$\ \textsuperscript{b} \\
$R_{500}$ & $\SI{1.43}{Mpc}$\ \textsuperscript{c} $\ \hat{=} \ \SI{20.7}{\arcmin}$\\
$R_{200}$ & $\SI{2.33}{Mpc}$\ \textsuperscript{c} $\ \hat{=} \ \SI{33.8}{\arcmin}$ \\
$R_{100}$ & $\SI{3.17}{Mpc}$\ \textsuperscript{d} $\ \hat{=} \ \SI{46.0}{\arcmin}$ \\
\bottomrule
\end{tabular}
\\
\tablefoot{\textsuperscript{a} \citet{MCXC}, \textsuperscript{b} \citet{Dehghan_2017}, \textsuperscript{c} \citet{Ettori_2019}, \textsuperscript{d} $R_{100} \approx 1.36R_{200}$ \citep{Reiprich_2013}}
\end{table}

The general properties of the cluster as used in this work are listed in \cref{tab:cluster_props}.
The assumed cosmology in this work is the flat $\Lambda$CDM, with the cosmological parameters $h = \num{0.7}$, $\Omega_{\text{m}} = \num{0.3}$, and $\Omega_{\Lambda} = \num{0.7}$. At the redshift of A3266, $\SI{1}{arcsec}$ corresponds to $\SI{1.15}{kpc}$. All uncertainties are given at the $\SI{68}{\%}$ confidence level.

\section{Data reduction}
\label{sec:data_reduction}
We used data from the eROSITA all-sky survey~1-5 (eRASS:5, skytiles 073153, 066153, 060153, 071150, 066150, 060150, 071147, 066147, and 061147) with the eROSITA pipeline processing version~c030. The data reduction was performed with the extended science analysis software system \texttt{eSASS} (\citealt{Brunner_2022, Merloni_2024}), version~240410.0.3.

All images were created in the $\num{0.3}$--$\SI{2.0}{\kilo\electronvolt}$ band. All seven telescope modules (TMs) were used, but the energy band was restricted to $\num{0.8}$--$\SI{2.0}{\kilo\electronvolt}$ for TMs 5 and 7. These two TMs are not equipped with an on-chip filter, and their CCD pixels can pile up leaked optical light from the Sun in the very soft energy band \citep{Predehl_2021}. We therefore applied the lower limit of $\SI{0.8}{keV}$ for them.

All the data reduction steps followed the procedure described in \cite{Reiprich_21}. We briefly recall the main steps below. First, all five surveys and seven TMs were combined and filtered per skytile for soft proton flares using a $\SI{3}{\sigma}$ upper cut on the light curves in the energy band $\num{5.0}$--$\SI{10.0}{\kilo\electronvolt}$. Second, The particle-induced background (PIB) was estimated from the observation data between $\num{6.7}$--$\SI{9}{\kilo\electronvolt}$ and converted into the lower-energy band using the hardness ratios of eROSITA filter-wheel-closed data. The PIB was then subtracted from all images and surface brightness profiles. Finally, the exposure correction accounted for varying exposure times and vignetting in the image and also adjusted for the use of different energy bands for the TMs with and without on-chip filter. We also used the exposure map to conduct a relative correction of the spatially varying Galactic absorption: Expected count rates were simulated for different hydrogen column densities $N_{\text{H,tot}}$ in the line of sight, and a correction factor with respect to the median $\overline{N}_{\text{H,tot}}$ in the field of view was multiplied with the exposure map. The $N_{\text{H,tot}}$ value is a combination of atomic hydrogen column density $N_{\text{HI}}$ from the HI4PI all-sky survey \citep{HI4PI} and molecular hydrogen column density $N_{\text{H2}}$ obtained with the method by \cite{Willingale_13}. A large-scale image of A3266 and its surroundings after applying these data reduction steps and a Gaussian smoothing of $\sigma = \SI{10}{pixels}\, (=\SI{40}{\arcsec})$ is shown in \cref{Fig:large-scale_overview}. All eROSITA images show the count rate of the combined seven~TMs in the energy band $\num{0.3}$--$\SI{2.0}{\kilo\electronvolt}$ in units of counts per second.

\begin{figure}[htbp]
    \centering
    \includegraphics[width=\hsize]{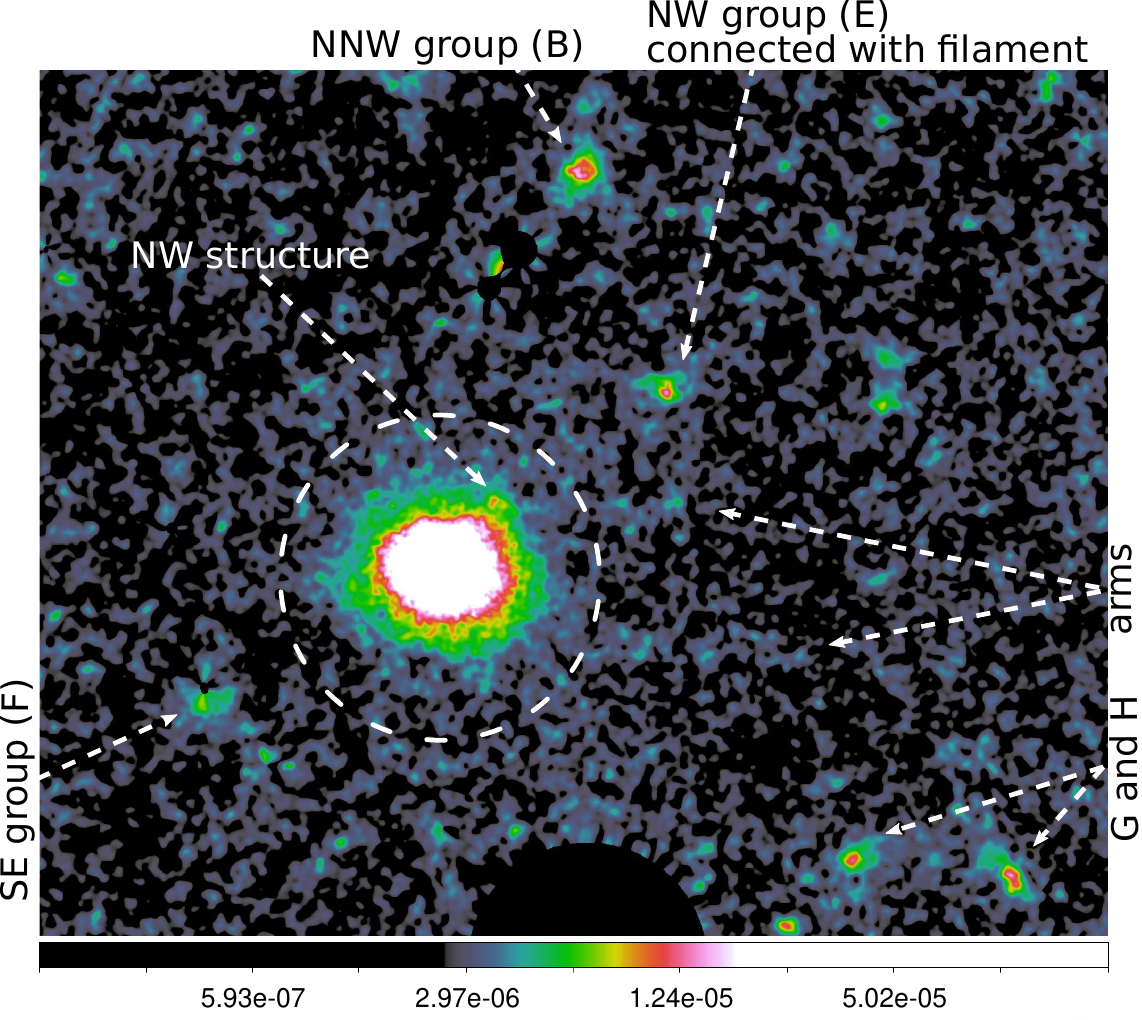}
    \caption{Data-reduced and adaptively smoothed eRASS:5 X-ray image of Abell 3266 and its outskirts in the $\num{0.3}$--$\SI{2.0}{\kilo\electronvolt}$ band. Fore- and background structures are excised, but groups within a $\SI{50}{Mpc}$ line-of-sight difference are kept. Overlaid is the $R_{200}$, and the annotations point out the most prominent features.
    }
    \label{Fig:SystemOverview}
\end{figure}

We removed the point sources for further image analysis through a combination of wavelet filtering, automated source detection, and manual cleaning and comparison to known catalogs. First, a wavelet-filtering algorithm based on \citet{Pacaud_2006}, optimized for Poisson-distributed data, was applied to the photon images. The algorithm is based on the à trous wavelet transform using a cubic B-spline wavelet, also known as the starlet transform (e.g., \citealt{Starck_Pierre_1998} for an astronomy example and \citealt{Starck_2015} for a detailed mathematical description). The wavelet decomposition separates the image into several spatial scales, allowing noise suppression by applying hard significance thresholds to the wavelet coefficients. A smoothed and denoised image is then iteratively reconstructed from the significant coefficients, taking the exposure time and PIB maps into account when determining the significance levels and during image reconstruction \citep{Faccioli_2018}.

Second, source detection was performed on the wavelet-filtered images using the software \texttt{Source Extractor} (SExtractor; \citealt{SExtractor}), following the procedure described in \citet{Pacaud_2006} and \citet{Faccioli_2018}. The resulting source catalog was manually cross-matched with the point sources and extended sources not belonging to the A3266 system identified in \citet{Sanders_2022}. Furthermore, foreground and background galaxy groups and clusters from the eROSITA DR1 catalog \citep{Bulbul_2024} were additionally included, with radii set to their $R_{500}$. The combined catalog was visually inspected and manually refined where necessary: a small number of apparent point sources not detected automatically were added, while spurious detections and extended sources in the fore- or background were removed. Moreover, the size of some extraction regions was adapted, for instance, in cases of overlapping and therefore imprecise detections. The final list of contaminating sources was used to generate a cheesemask that was applied to the original image to exclude the contaminating emission. This source list was also used to exclude the regions during the spectral fitting. \cref{Fig:SystemOverview} shows the final result and was additionally adaptively smoothed with the \texttt{asmooth} task from the science analysis software \texttt{SAS}\footnote{\url{https://www.cosmos.esa.int/web/xmm-newton/sas}}, version~20.0.0, using a minimum scale, maximum scale, and signal-to-noise ratio threshold of 0, 10, and 10, respectively.

Another image-processing method for visualization purposes only was a second wavelet filtering after contaminating sources were removed. It again suppressed Poisson noise and enhanced the emission features on very large scales. This is shown in \cref{Fig:wavelet_SB_defs} and in \cref{Fig:slow_compare} in the left panel. Wavelet filtering can produce ringing or horseshoe-shaped artifacts that appear as negative rings or bowls around very bright circular or elliptical sources (e.g., \citealt{Starck_2015}; see an example in \citealt{Dietl_2024}, Appendix A). In our case, the point sources were already removed and therefore did not contribute to these artifacts. However, the cluster core of A3266 is approximately two orders of magnitude brighter than its outskirts. To prevent artifacts around A3266, the core was excluded from the image, which reduced the overall dynamical range.

\section{Analysis and results}
\label{sec:analysis}

\subsection{Imaging analysis}
\label{subsec:imaging_analysis}
\begin{figure}[htbp]
    \centering
    \includegraphics[width=\hsize]{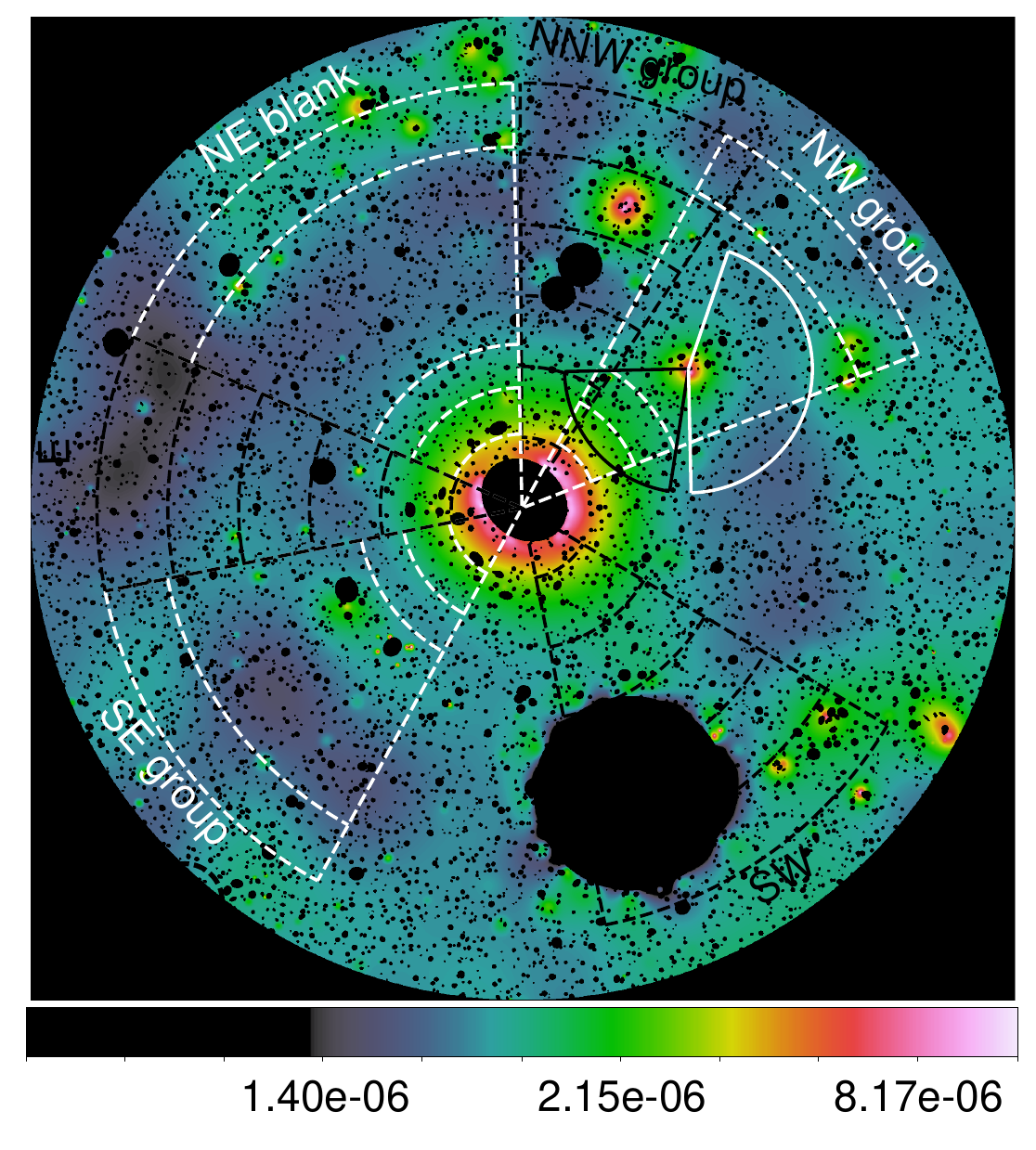}
    \caption{Wavelet-filtered image after source removal, cut to $3R_{100}$. The NW and NNW groups and the filament connecting the NW group to A3266 are clearly visible. The center of A3266 is excised to avoid wavelet-filtering artifacts; the largest excised sources correspond to the sources J, L, and M (see~\cref{Fig:large-scale_overview}). The overlaid sectors define the surface brightness profiles. The dashed black sectors show annuli at $[20,40,60,80,100, and120]\si{\arcmin}$, and the dashed white sectors mark $R_{500}$, $R_{200}$, $R_{100}$, $3R_{200}$, and $\SI{120}{\arcmin}$. The solid pie regions were used to extract the cluster-facing and cluster-opposite profiles of the NW group.}
    \label{Fig:wavelet_SB_defs}
\end{figure}

\cref{Fig:SystemOverview} reveals multiple large-scale features around A3266, including the NW structure, two arms extending westward, and multiple groups with similar redshift. We performed a quantitative surface brightness analysis in various directions of the cluster. The sectors were chosen to follow the main structures that are visible in the image. The configuration of sectors we used is shown in \cref{Fig:wavelet_SB_defs} (dashed sectors). The resulting exposure- and relative-absorption-corrected and PIB-subtracted surface brightness profiles for the azimuthally averaged profile and the NNW, SE, and NW sectors, computed with \texttt{pyproffit} \citep{Eckert_2020}, are shown in \cref{Fig:SB1} to \cref{Fig:SB2a}. The profiles of the remaining sectors are displayed in the appendix in \cref{Fig:SB3}. Our interest was to study the outskirts of A3266. The profiles are therefore shown with a linear scale on the $x$-axis, and the $y$-axis was cut to enhance the visibility of the low surface-surface brightness changes. Thus, the core of A3266 cannot be seen in these profiles (we refer to the literature in the Introduction for detailed studies of the X-ray morphology in the core).

An azimuthally averaged profile was extracted for a baseline profile that we used for comparison with the other sectors. All groups were excised out to their $R_{200}$ for this profile in order to only show the ICM emission of A3266. A double-$\beta$ model with an additive constant for the unresolved cosmic X-ray background (CXB) was fitted to the data. The sector profiles were extracted from the data, including the surrounding groups. The profiles are discussed in \cref{sec:discussion}.

A region of enhanced emission lies between A3266 and the NW group (see~\cref{Fig:SystemOverview} and \cref{Fig:SB2a}), and the NW group and its connection to the main cluster is therefore particularly interesting.
The profile between the two $R_{200}$ radii of A3266 and the NW group has a $\SI{13(2)}{\%}$ enhanced emission with $\SI{5.4}{\sigma}$ compared to the CXB level.
To assess whether this excess can be explained by the superposition of the cluster and the group emission, we constructed a model for the expected surface brightness. For the NW group, we extracted a full azimuthal profile, which can be well described by a single $\beta$-model. For A3266, we modeled the surface brightness using the profile extracted in the NW sector, which points toward the group, and fitted a double-$\beta$ model. We adopted the NW sector profile because A3266 shows an elongated morphology in this direction and the model is meant to represent the outskirts of A3266 in this direction in particular. The combined model is displayed in \cref{Fig:SB2a}. Comparing the measured profile to this model, we found a remaining surface-brightness enhancement of $\SI{8(2)}{\percent}$ with a significance of $\SI{3.6}{\sigma}$, revealing a filament structure between A3266 and the NW group.

We also defined two sectors for the NW group, one cluster-facing and the other cluster-opposite (solid line sectors in \cref{Fig:wavelet_SB_defs}). The cluster-opposite profile follows the full-azimuthal profile and its single $\beta$-model fit very well (\cref{Fig:SB_group} upper panel), while the cluster-facing profile exhibits a sharp drop in the inner part of the group. With the \texttt{BknPow} model in \texttt{pyproffit}\footnote{\url{https://pyproffit.readthedocs.io/en/latest/Discontinuities.html\#Modeling-the-brightness-profile}}, a broken power-law fit determines a density jump of $\num{2.8(7)}$ at $r_{\rm f} = \SI{1.83(2)}{\arcmin}$ (\cref{Fig:SB_group} lower panel).

The significance of the arm features (see~\cref{Fig:SystemOverview}) was evaluated by selecting a rectangular region with a width of \SI{6}{arcmin} for each arm, located between $R_{200}$ and $2R_{200}$, and comparing its emission to that of two adjacent regions with identical sizes placed directly next to the respective arm. These comparison regions exhibited consistent surface brightness levels, and we therefore adopted their mean value as the reference for each arm. This analysis yielded a significance of $\SI{4.6}{\sigma}$ for the northern arm and of $\SI{4.9}{\sigma}$ for the southern arm.

\begin{figure}[htbp]
    \centering
    \includegraphics[width=0.9\hsize]{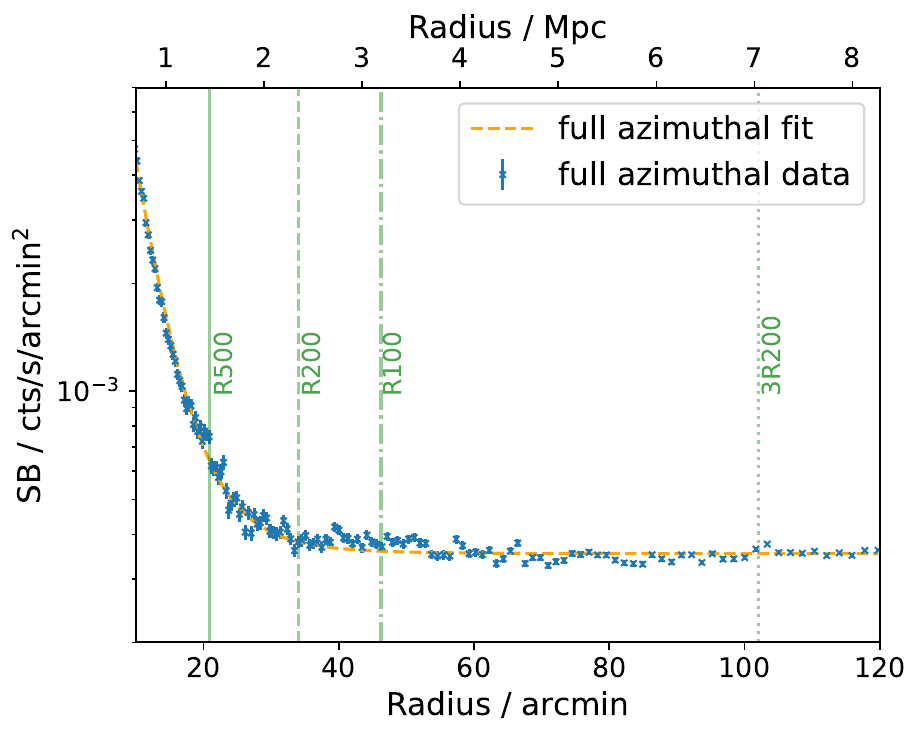}
    \caption{Azimuthally averaged profile of A3266 with the double-$\beta$ model fit; all surrounding groups are excised.}
    \label{Fig:SB1}
\end{figure}

\begin{figure}[htbp]
    \centering
    \includegraphics[width=0.9\hsize]{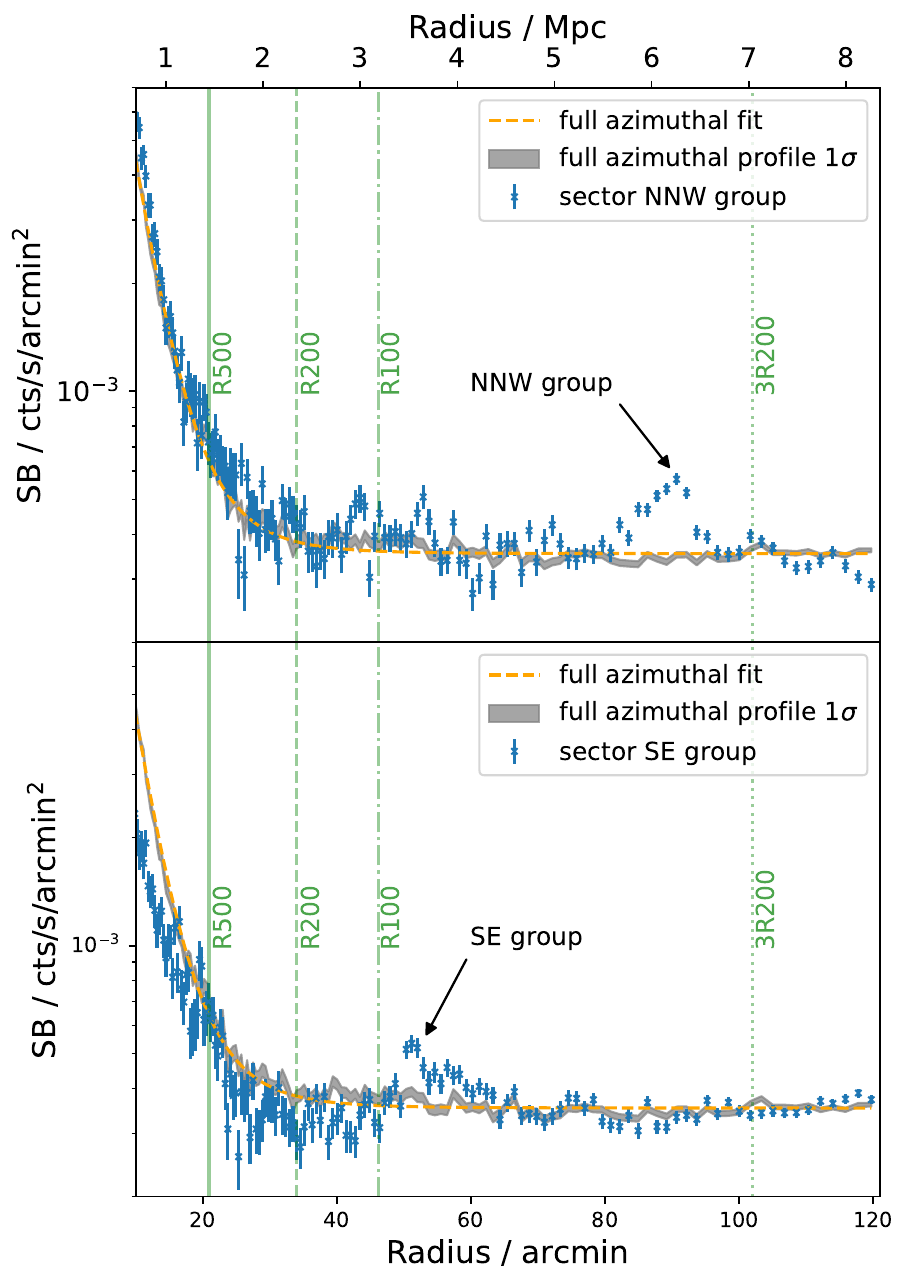}
    \caption{Example sector profiles of A3266. Surrounding groups are not excised. The azimuthally averaged fit and profile without the surrounding groups is shown for comparison.
    }
    \label{Fig:SB2}
\end{figure}

\begin{figure}[htbp]
    \centering
    \includegraphics[width=0.9\hsize]{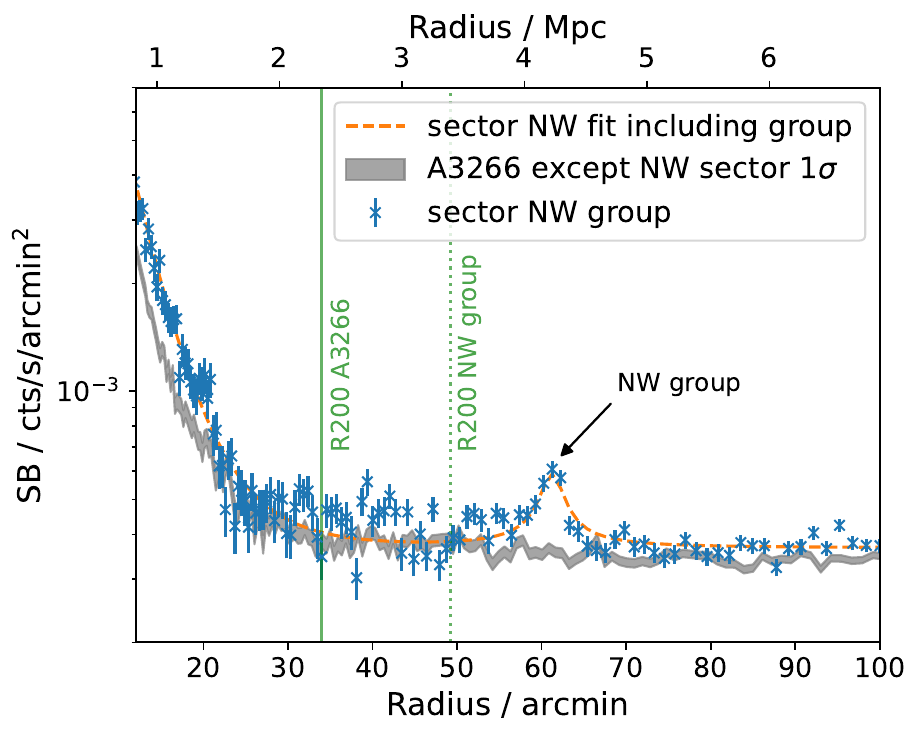}
    \caption{NW sector profile of A3266. The NW group is not excised. This plot studies the excess emission compared to a model accounting for the NW elongation of A3266 and including the NW group (see text for details).
    }
    \label{Fig:SB2a}
\end{figure}

\begin{figure}[htbp]
    \centering
    \includegraphics[width=0.9\hsize]{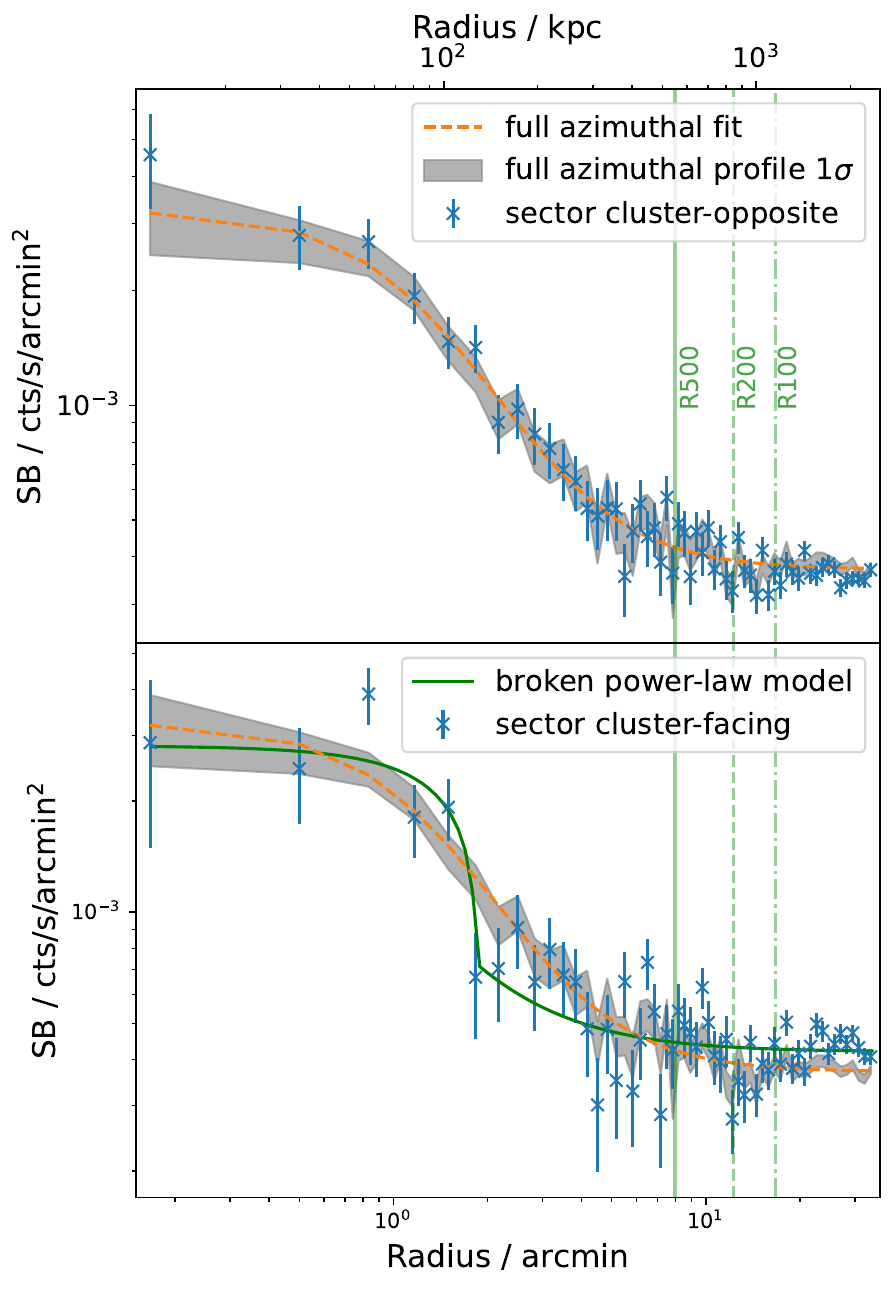}
    \caption{Cluster-opposite and cluster-facing sector profiles of the NW group. The latter profile features a surface brightness discontinuity, fitted by a broken power-law model. The labels that are not repeated in the second panel are the same as in the upper panel.
    }
    \label{Fig:SB_group}
\end{figure}

\subsection{Spectral analysis}
\label{subsec:spectral_analysis}

To characterize the gas density, temperature, and metallicity of the NW group and of the filament between A3266 and the NW group, we performed a spectral analysis following the method described in \citet{Veronica_2024}. Only the main steps are summarized below.

\begin{figure}[htbp]
    \centering
    \includegraphics[width=0.8\hsize]{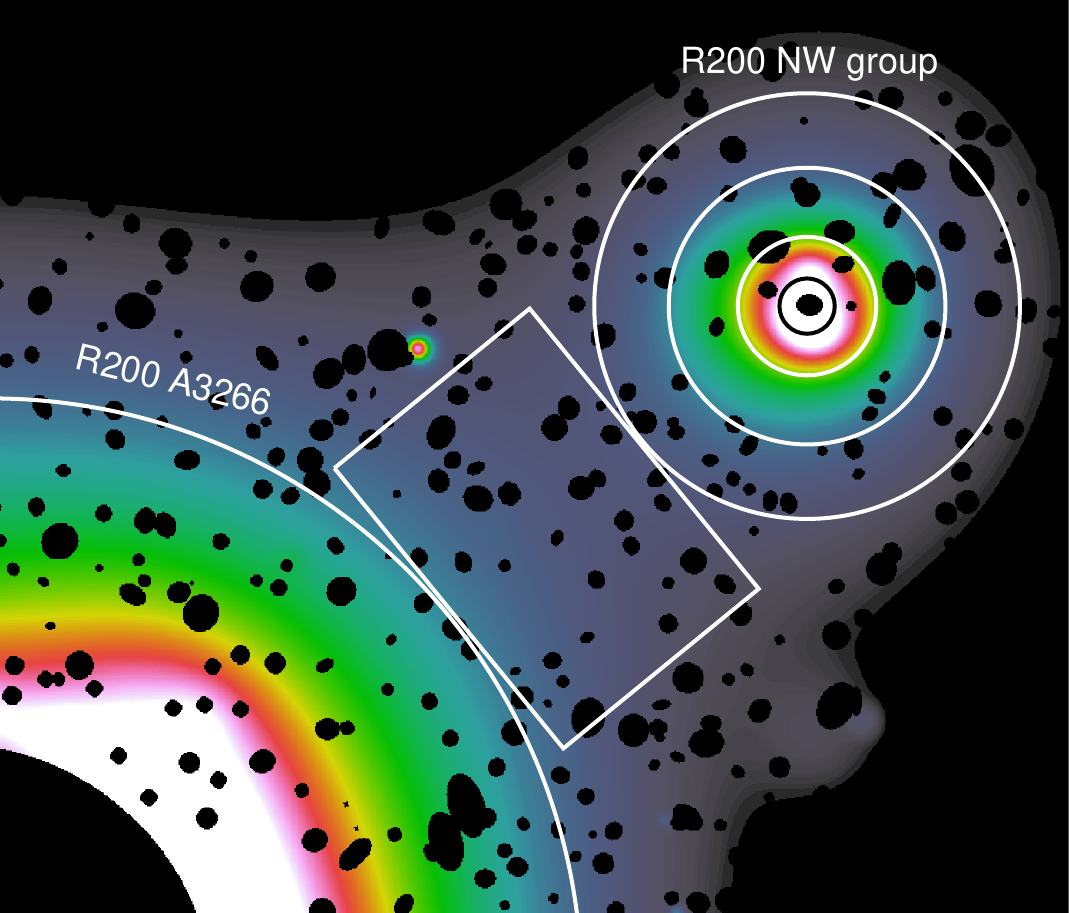}
    \caption{Configuration of spectral regions for \cref{Fig:spectra}. The annuli used for the NW group are $<0.2R_{500}$, $0.2$--$0.5R_{500}$, $0.5R_{500}$--$1R_{500}$, and $R_{500}$--$R_{200}$. The first annulus is colored differently for better visibility. The filament box is located between the $R_{200}$ radii of A3266 and the NW group and is $\SI{1.5}{Mpc}$ wide.}
    \label{Fig:spectral_regions}
\end{figure}

Spectra, ancillary response files (ARFs), and response matrix files (RMFs) for the source and background regions were extracted using the \texttt{eSASS} task \texttt{srctool}. The extraction region for the filament was defined by a rectangular region, and the NW group was divided into four annuli: $0$–$0.2R_{500}$, $0.2R_{500}$–$0.5R_{500}$, $0.5R_{500}$–$R_{500}$, and $R_{500}$–$R_{200}$. The locations of these regions are shown in \cref{Fig:spectral_regions}. The background was extracted from a circular region with a radius of $\SI{20}{\arcmin}$ in the \texttt{NE blank} sector, located between $R_{100}$ and $3R_{200}$ of A3266.

All seven TMs were used in the analysis. TMs equipped with on-chip filters were fitted in the $\num{0.3}$–$\SI{9.0}{\kilo\electronvolt}$ energy range, while the remaining two TMs were restricted to $\num{0.8}$–$\SI{9.0}{\kilo\electronvolt}$. The spectral fitting was performed with \texttt{XSPEC} \citep{xspec}, using a model that includes the particle-induced background (PIB; \citealt{Yeung_2023}), the unabsorbed local hot bubble (LHB), absorbed Milky Way halo (MWH), unresolved cosmic X-ray background (CXB), and an absorbed thermal source component. The total hydrogen column density was adopted from the $N_{\mathrm{H,tot}}$ map described in \cref{sec:data_reduction} using the median value within each extraction region for the corresponding absorption component.

The background spectra were fitted first using a fixed temperature and metallicity for the LHB of $\SI{0.1}{keV}$ and $\SI{1}{Z_\odot}$ (adopted from \citealt{Veronica_2024}). The temperature and metallicity of the MWH were allowed to vary and constrained to $T_{\text{MWH}} = \SI{0.209(3)}{keV}$ and $Z_{\text{MWH}} = \SI{0.5(1)}{Z_\odot}$. The CXB was modeled as an absorbed power law. Its photon index was allowed to vary during the background fit and was constrained to $\Gamma=\num{1.50(2)}$, consistent with previous measurements of the unresolved CXB (e.g.,\citealt{Cappelluti_2017} and \citealt{Yeung_2024}).

Subsequently, the source spectra were fitted using the background parameters constrained in the previous step. All normalizations were thawed in the fit. Since the best-fit background parameters can vary between different sky positions and differ slightly from those adopted in previous X-ray studies, we additionally repeated all fits using $T_{\text{MWH}} = \SI{0.25}{keV}$, $Z_{\text{MWH}} = \SI{1.0}{Z_\odot}$, and $\Gamma = \num{1.46}$, which are the values used in \citet{Veronica_2024}. The resulting source parameters remain consistent within less than $\SI{1}{\sigma}$. We used the metallicity tables by \citet{Asplund_2009} in the plasma emission and absorption models. 

\begin{figure}[htbp]
    \centering
    \includegraphics[width=\hsize]{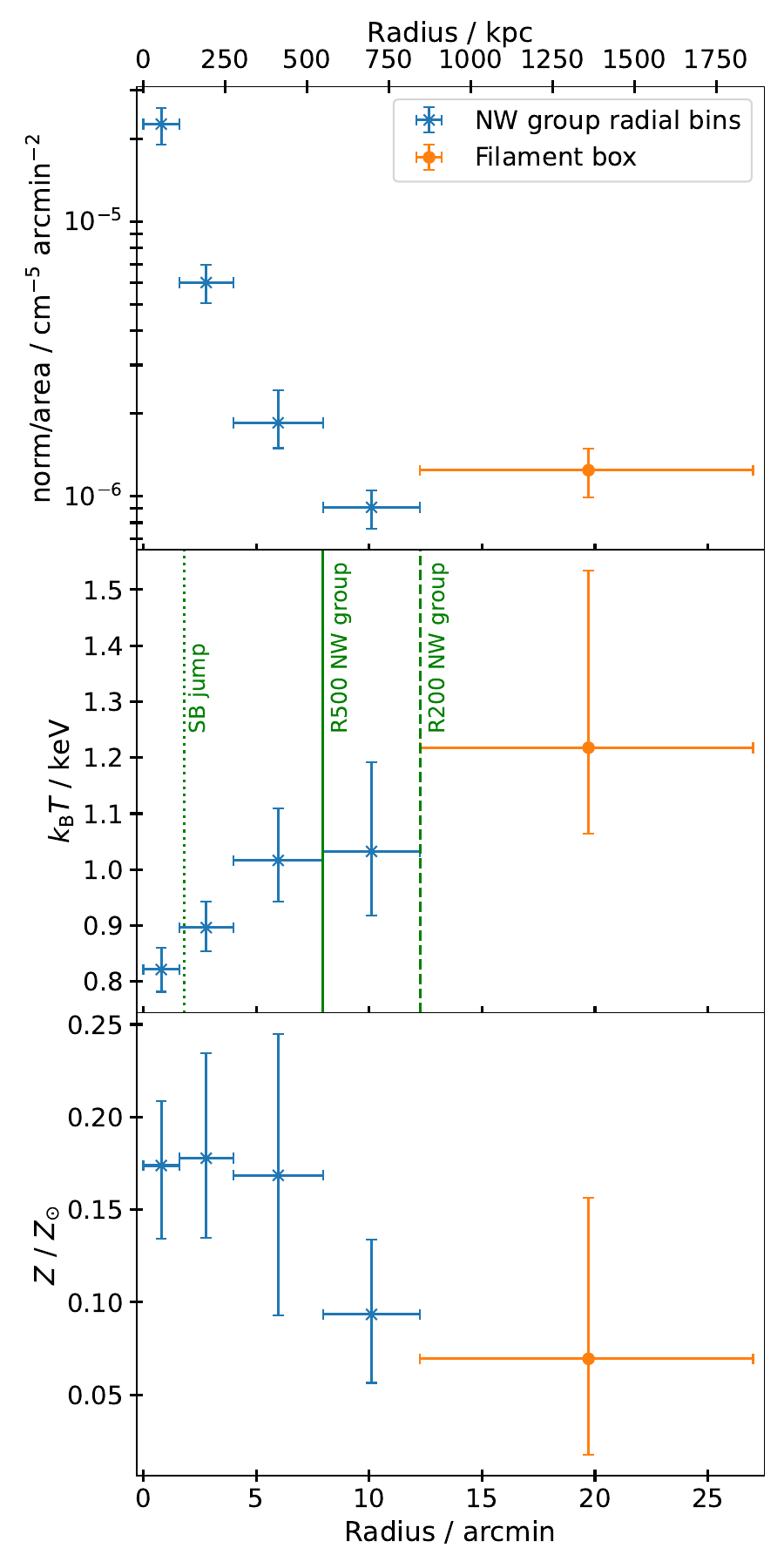}
    \caption{Results for normalization, temperature, and metallicity of the NW group. The spectral regions are defined in \cref{Fig:spectral_regions}. The error bars represent the fitting uncertainties at the $\SI{68}{\%}$ confidence level.}
    \label{Fig:spectra}
\end{figure}

\cref{Fig:spectra} presents the results of the spectral analysis, and \cref{tab:spectral_results} provides a summary of all fitted parameters. \cref{fig:TM1_spectrum} shows an example of a fitted TM1 spectrum for the filament region. We also attempted a two-temperature fit on the filament region because the gas might consist of a cluster outskirt component and an inflowing gas component, but the normalization of this additional temperature component converged to 0.

For A3266 itself, we fitted a temperature of $\num{4.2}_{-1.7}^{+3.3}\,\mathrm{keV}$ in the range $R_{500}$--$R_{200}$ in the NW sector (in the direction of the filament) and a temperature of $\num{3.8}_{-1.4}^{+3.7}\,\mathrm{keV}$ in the the same radial range, but in the \texttt{NE blank} sector (in a direction without a filament or group). The NW fit constrained the abundance to $Z = \SI{0.34(29)}{Z_\odot}$, and the NE fit only converged when we fixed the abundance. No significant temperature change was observed when we fixed the abundance to $Z=\SI{0.2}{Z_\odot}$ or $Z=\SI{0.3}{Z_\odot}$.

For the NW group, the annuli were additionally divided into cluster-facing and cluster-opposite sectors, analogous to the surface brightness analysis. The resulting temperature and metallicity measurements are consistent within their uncertainties, and the spectral fits do not reveal significant differences compared to those obtained from the full annuli. Given the relatively large radial bins and the current data quality, we cannot exclude a temperature structure associated with the observed surface brightness jump. However, no statistically significant temperature asymmetry is detected in the present data.
Finally, the redshift was allowed to vary in a fit to the spectrum extracted within $R_{500}$, yielding $z_{\rm NW group} = 0.06^{+0.01}_{-0.02}$.

\subsection{Filament gas density}
\label{subsec:filament_gas}
We used the derived normalization from the \texttt{apec} fit component to calculate the electron density $n_{\rm e}$ within in the extraction region displayed in \cref{Fig:spectral_regions}. We assumed a cylindrical morphology to convert the emission measure, analogous to \citealt{Veronica_2024},
\begin{equation}
\begin{aligned}
n_{\rm e} = \Bigl[&
\SI{1.52e-10}{\per\centi\meter} \times \textnormal{norm} \times (1+z)^2 \\
&\times \left(\frac{D_{\rm A}}{\rm Mpc}\right)^2
\times \left(\frac{r}{\rm Mpc}\right)^{-2}
\times \left(\frac{h}{\rm Mpc}\right)^{-1}
\Bigg]^{\frac{1}{2}} \, ,
\end{aligned}
\end{equation}
where $h$ is the height of the cylinder, that is, the length of the filament, and $r$ is the radius of the cylinder, that is, half the width and depth of the filament. The first factor is a conversion factor from the \texttt{xspec} normalization, and we assumed $n_{\rm H} = \frac{n_{\rm e}}{1.17}$. The baryon overdensity corresponds to
\begin{align}
    \delta_{\rm b} := \frac{\rho_{\rm gas}}{\overline{\rho_{\rm b}}(z)} = \frac{\mu_{\rm e}n_{\rm e}m_{\rm p}}{\Omega_{\rm b}(z) \rho_{\rm crit}(z)} \ ,
\end{align}
where $\rho_{\rm crit}(z)$ and $\Omega_{\rm b}(z)$ are the critical density of the Universe and the cosmological baryon density parameter at the redshift of the filament. $m_{\rm p}$ is the proton mass, and $\mu_{\rm e} = 1.153$ the mean molecular weight per free electron.

The spectral extraction region box (see~\cref{Fig:spectral_regions}) was placed between the $R_{200}$ radii of A3266 and the NW group, which corresponds to $h = \SI{1}{Mpc}$, and its width was chosen to capture the part of roughly constant surface brightness between cluster and group, which led to $r = \SI{0.72}{Mpc}$. Furthermore, we accounted for flux lost due to excised point sources by computing the fraction $f$ of emission measure retained using a projected cylindrical geometry. This yielded a factor of $f=0.88$, corresponding to a \SI{6.7}{\percent} increase in the derived electron densities.
The redshift $z_{\rm A3266} = \num{0.0596}$ results in $n_{\rm e} = \SI{8.3(8)e-5}{\per\cubic\centi\meter}$ and $\delta_{\rm b} = \num{320(30)}$. The uncertainties reported here are only the statistical uncertainties, and the results were calculated for the case that the filament lies in the plane of the sky. The effect of this assumption and an estimation of the systematic uncertainties is discussed in \cref{sec:discussion}.

\section{Discussion}
\label{sec:discussion}
A3266 is a massive nearby galaxy cluster at the southeastern edge of the HRSC. It is embedded in a complex large-scale environment and surrounded by several galaxy groups and clusters at comparable redshifts (\cref{Fig:large-scale_overview}). The systems labeled A, B, and C were identified as members of a common supercluster by \citet{Liu_2024}; E and F form a different supercluster, but at a redshift consistent with the first. We argue that all the neighboring groups D, E, F, G, H, and I might be part of the same structure. Their redshift differences correspond to line-of-sight separations smaller than \SI{50}{Mpc}, even when we do not account for the velocity dispersion of A3266, which alone is sufficient to explain the observed offsets in redshift. The friends-of-friends algorithm employed by \citet{Liu_2024} might not associate these systems with the same supercluster because it is highly sensitive to redshift differences. This sensitivity is particularly relevant because many of the smaller groups show redshift separations of about $\Delta z \approx 0.01$, comparable to their photometric redshift uncertainties (\cref{tab:group_props}), making it difficult to interpret these separations as physically meaningful.

\subsection{Surface brightness structures}
\label{subsec:SBstructs}
\cref{Fig:SystemOverview} shows the X-ray image of A3266 after full data reduction, covering a region extending beyond $3R_{200}$. The cluster displays an east–west elongation with a northwestern extension at approximately $R_{500}$, labeled the NW structure by \citet{Sanders_2022}. This structure corresponds to a previously identified subgroup (structure~4 in \citealt{Dehghan_2017} and region~9 in \citealt{Gatuzz_2024}), which may have undergone a passage through the cluster core \citep{Sanders_2022}. At larger radii, two narrow arm-like features extend toward the west ($\SI{4.6}{\sigma}$ and $\SI{4.9}{\sigma}$ significance). They reach roughly out to $R_{100}$ and appear to be significantly narrower and less diffuse than the filamentary structure connecting the cluster to the NW group. While diffuse cluster emission is detected out to similar radii in other directions, these features appear to be localized surface brightness enhancements. Such structures in cluster outskirts might arise from anisotropic gas distributions related to ongoing accretion or stripped gas from infalling substructures.

In all directions except the southeast, the emission can be traced out to the virial radius, assuming $R_{\textnormal{virial}} \approx R_{100}$. Toward the northwest, the emission continues beyond the NW structure and forms a filament connecting the cluster to the NW group, followed by a faint tail extending to ${\sim}3R_{200}$. In contrast, the southeast shows no detectable connection to the SE group, with emission dropping even below the CXB level. Possible excess emission is seen toward the NNW group, although contamination from the projected background clusters L and M cannot be excluded, even though they were excised. A similar situation occurs in the southwest, where the foreground galaxy cluster J was excised, but residual contamination might still contribute to the apparent excess emission. The image also includes groups G and H in the lower right corner. These groups are connected by an X-ray bridge.

The most prominent features identified above were quantitatively analyzed using surface brightness profiles. The azimuthally averaged profile shown in \cref{Fig:SB1} is well described by a double-$\beta$ model and remains above the sky background out to $R \approx \SI{55}{\arcmin} > R_{100}$ with a significance of $\SI{4}{\sigma}$. In addition, the profile exhibits a clear jump at $R_{500}$, which is primarily driven by the northwestern direction. All identified clumps and galaxy groups were excised prior to extracting this profile and therefore did not contribute to the measured emission.

\cref{Fig:SB2} presents surface brightness profiles extracted in different sectors, where groups and clumps were not removed. In the upper panel, the NNW group is clearly detected. Between A3266 and the NNW group, a wave-like pattern is visible in the profile. While this feature might be associated with gas sloshing, no corresponding evidence is seen in the X-ray image. Moreover, beyond $R_{100}$, residual contamination from the excised background clusters cannot be excluded, making it unclear whether the excess emission originates from a filamentary connection to the NNW group or from projected emission along the line of sight.

The lower panel highlights the excess emission associated with the SE group, which appears to be asymmetric and extends farther away from A3266 than toward it. The emission between the cluster and the group lies not only below the azimuthally averaged cluster profile, but also below the sky background level. In addition, the emission within $R_{500}$ in this sector remains below the cluster average.

\cref{Fig:SB2a} shows the surface brightness profile in the northwestern direction. Here, the emission lies systematically above the cluster average and displays a sharp drop at $R_{500}$, caused by the NW structure. Excess emission is clearly detected between A3266 and the NW group, and the emission at larger radii beyond the group also remains elevated with respect to the azimuthally averaged profile. The excess emission between the $R_{200}$ radii of the cluster and the NW group amounts to $\SI{13(2)}{\percent}$ at a significance of $\SI{5.4}{\sigma}$ relative to the cluster average. However, since A3266 is strongly elongated in this direction, a comparison with the azimuthally averaged profile alone does not provide a conservative estimate, and the outskirts of the cluster and the NW group may partially overlap. To account for this, we fitted a separate double-$\beta$ model using only the data from the northwestern sector and combined it with a single-$\beta$ model for the NW group, thereby explicitly modeling the elongated morphology of A3266 and the overlapping cluster and group emission. The observed surface brightness between the two systems exceeds this model at a significance of $\SI{3.6}{\sigma}$, corresponding to an additional $\SI{8(2)}{\percent}$ enhancement, which we interpret as emission from a filament connecting A3266 and the NW group.

The profiles of the remaining sectors NE, E, and SW show less significant deviations from the azimuthally averaged profile and can be found in \cref{sec:app_SB_profs}.

While fitting a surface brightness model to the NW group, we identified a pronounced asymmetry between the sector facing the cluster and the sector pointing away from it. As shown in \cref{Fig:SB_group}, the cluster-opposite sector closely follows the azimuthally averaged surface brightness profile and its best-fitting $\beta$-model, whereas the cluster-facing sector exhibits a sharp drop. This feature corresponds to a density jump of $\num{2.8(7)}$, as determined from a broken power-law fit. Since such a discontinuity might either be caused by a cold front or a shock front, additional constraints from the temperature structure are required to distinguish between these scenarios. This is discussed below.

\subsection{Comparison to the distribution of member galaxies}
\label{subsec:galaxies}
\begin{figure}[htbp]
    \centering
    \includegraphics[width=\hsize]{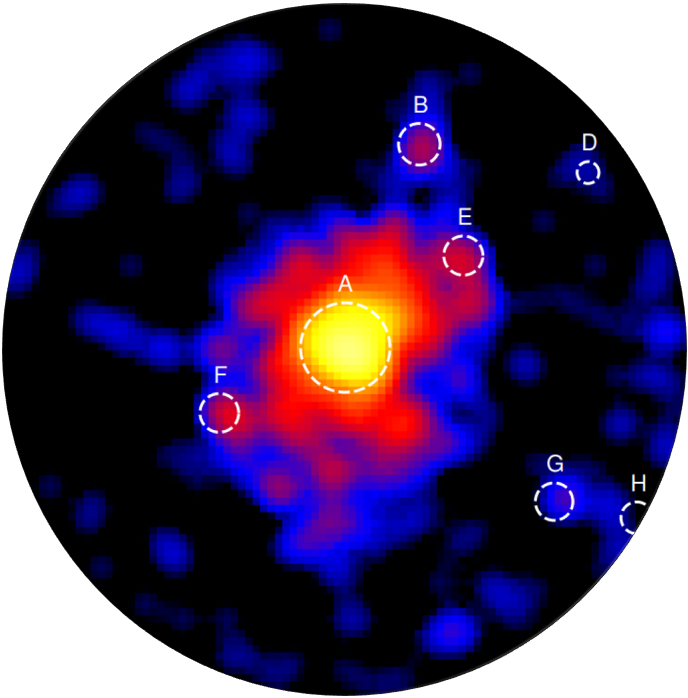}
    \caption{Galaxy density map of galaxies extracted from NED with a redshift within $3v_{\rm disp}$. The image section is the same as in \cref{Fig:wavelet_SB_defs}. Galaxy groups from \citet{Bulbul_2024} are overlaid (see~\cref{Fig:large-scale_overview}). We observe galaxies far beyond $R_{500}$, especially in NW direction, and extending towards B, E, and F.}
    \label{Fig:galaxy_map}
\end{figure}

Galaxies with spectroscopic and photometric redshifts within $z_{\rm A3266} \pm 3v_{\rm disp}$, that is, $z \in [0.0462, 0.0730]$, were retrieved from NED\footnote{\url{https://ned.ipac.caltech.edu/}}. We used the velocity dispersion from \citealt{Dehghan_2017} (see~\cref{tab:cluster_props}). The wide redshift range was not used to represent virialized motions within the cluster, but relative motions of multiple infalling structures. \cref{Fig:galaxy_map} presents a map of the retrieved galaxies using the same image section as \cref{Fig:wavelet_SB_defs}.

The map consists of 693 spectroscopic redshift measurements and 1382 photometric measurements. When only galaxies with spectroscopic redshifts are used, most features in the map remain the same, but with fewer galaxies. Notable features that are no longer present are the extension toward the south, the overdensities of groups D and F, and the connection between G and H. Only one spectroscopic galaxy is available for groups D and F.

The projected distribution of member galaxies extends out to $\approx \SI{50}{\arcmin}$, comparable to the radius over which the ICM can be traced in X-rays. All associated groups are located at positions with an enhanced galaxy density. However, for groups D and F, this is not spectroscopically confirmed because there is only one spectroscopic redshift measurement for these groups.

At approximately $R_{500}$ of A3266, the galaxy distribution closely follows the X-ray morphology, with a higher density toward the northwest than toward the southwest. At larger radii, some structures show a clear correspondence between galaxies and X-ray emission, while others do not. The NW and NNW groups are connected to A3266 by galaxy overdensities, with the connection to the NW group being more prominent, in agreement with the X-ray morphology. This correspondence additionally supports the hypothesis that a filamentary structure connects A3266 to these groups. Galaxy overdensities aligned with enhanced X-ray emission suggest that galaxies and hot gas trace the same large-scale structure in these directions. This consistency agrees with expectations for filaments feeding material into the cluster.

In contrast, toward the southeast, a galaxy overdensity appears to connect the X-ray-detected SE group to the main cluster, while no corresponding X-ray emission is observed along the apparent galaxy connection. Because this feature was identified using photometric redshifts, the overdensity might be enhanced by projection effects and needs not represent a physical bridge between the SE group and A3266. Spectroscopic redshifts are required to determine whether the galaxies in this apparent connection are part of a common structure. If the overdensity were tracing a real connection, the lack of detectable X-ray emission might indicate that the diffuse gas in this region has not been heated sufficiently (e.g., through compression during infall) to reach temperatures detectable with eROSITA. Furthermore, the X-ray surface brightness in this sector between the SE group and the main cluster falls below the CXB level, which might suggest that hot gas has been substantially disturbed, displaced, or removed, potentially as a consequence of past merger activity in A3266, while the galaxy distribution continues to trace the underlying large-scale structure.

Another galaxy overdensity toward the southwest extends in the direction of groups G and H, which is not as prominent in X-rays. The associated X-ray emission may be too faint to be robustly detected or may be affected by residual contamination from the foreground cluster projected along this line of sight.

\subsection{Comparison to the SLOW simulation}

\begin{figure}[htbp]
    \centering
    \includegraphics[width=\hsize]{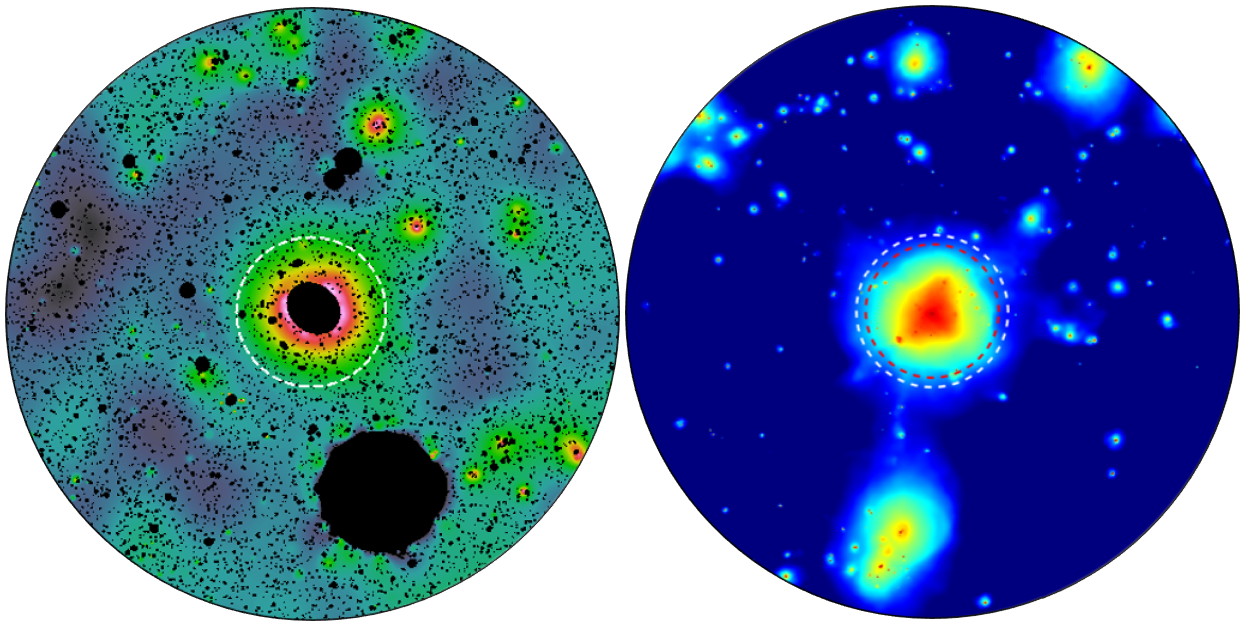}
    \caption{Left: Wavelet-filtered eRASS:5 X-ray image with overlaid $R_{200}$. Right: Same cluster with a matched image section and matched rotation in the constrained simulation \textit{SLOW} \citep{Dolag_2023}. The same $R_{200}$ as in the left panel is overlaid in white, as is $R_{200}$ from the simulation in red.}
    \label{Fig:slow_compare}
\end{figure}

\cref{Fig:slow_compare} shows the eRASS:5 wavelet-filtered X-ray image and the corresponding region from the constrained hydrodynamical simulation SLOW (e.g., \citealt{Dolag_2023,Hernandez_2024}) with a projection depth of $\pm\SI{15}{Mpc}$ around the cluster redshift. SLOW was designed to reproduce the structure of the nearby Universe by matching observed large-scale density modes. A3266 lies close to the high-redshift boundary of the SLOW volume, so that an exact match to the observed sky is not expected (Appendix I in \citealt{Seidel_2025}). Nevertheless, the simulated cluster shows a similar extent around $R_{200}$ and a qualitatively comparable large-scale environment, as described below.

In the northwest, the simulation contains the NW group connected to the cluster by a filament and also the more distant NNW group near the image boundary that is less directly connected. One of the western arms visible in the X-ray data (see~\cref{Fig:SystemOverview}) is also broadly reproduced in the simulation. However, the SE group lacks a clear counterpart. Instead, the simulation shows an additional double-core group merging from the south that is not detected in the X-ray image. Although this group seems to be at the same position as the foreground galaxy cluster J (see~\cref{Fig:large-scale_overview}), it is a different object since the projection depth of the image does not include cluster J. Instead, it might represent the same type of accretion event as the SE group, but from a slightly different direction in the simulation. It is also possible that this double-core group in the SLOW simulation has already merged with A3266, given its known complex structures in the core due to recent mergers. In this context, the apparent mismatch between the overdensity of galaxies toward the SE and the absence of a corresponding X-ray excess might be tentatively interpreted as evidence that a fraction of the galaxy population might be spatially segregated from the X-ray–emitting gas caused by the past merger, as suggested in \cref{subsec:galaxies}. 
Toward the northeast, diffuse X-ray emission corresponds to several clumps in the simulation near the image boundary, although the redshift of the emission in the eROSITA data is unknown.

Overall, SLOW depicts an actively accreting cluster with a similar number of surrounding groups and filamentary connections, although the detailed positions and sizes of individual structures differ from the observations.

\subsection{Cool-core NW group with a cold front}
The spectral analysis (\cref{Fig:spectra}) indicates that the NW group hosts a cool core. The temperature rises within the innermost $\approx \SI{300}{kpc}$ ($\approx 0.5R_{500}$) and remains approximately constant at $\approx \SI{1.0}{keV}$ at larger radii. The filament region is slightly hotter, with a temperature of $\num{1.2}_{-0.2}^{+0.3}\,$keV.
Instead of the cool-core scenario, the central temperature measurement might also reflect the average temperature of the group, while the outer regions toward A3266 are dominated by shock-heated gas, leading to higher measured temperatures. However, since we divided the spectral annuli into cluster-facing and cluster-opposite sectors and observed a consistent increase in temperature in both directions out to $R_{200}$, the cool-core interpretation is more consistent with the data.

We did not observe a clear temperature jump at the location of the surface brightness discontinuity, but the spectral bins required to achieve a sufficiently high signal-to-noise ratio are substantially broader than the width of the edge, so a sharp temperature change is not expected to be resolved. Nevertheless, the temperature is clearly lower toward the group center, supporting the interpretation that the observed density jump corresponds to a cold front. In contrast, a shock front would require higher temperatures on the denser side of the discontinuity, which is not observed. This interpretation is consistent with a scenario in which a cool-core group moves through the hotter filament gas toward the main cluster, producing a contact discontinuity where the leading edge is shaped by ram-pressure confinement.

The metallicity of the NW group is consistent with a radially flat profile from the center out to $R_{500}$, but lies at the lower end overall, with a best-fitting value of $Z = \SI{0.17(8)}{Z_\odot}$. For comparison, \citet{Mernier_2017} report typical metallicities of $\approx \SIrange[range-phrase={\text{\textendash}}]{0.6}{1.0}{Z_\odot}$ in the centers of galaxy groups and $\approx \SIrange[range-phrase={\text{\textendash}}]{0.15}{0.4}{Z_\odot}$ at $R_{500}$.

\subsection{Filament gas properties}
The temperature of the filament gas ($\num{1.2}_{-0.2}^{+0.3}\,$keV) exceeds typical expectations for warm-hot intergalactic medium (WHIM) filaments (e.g. $<\SI{0.86}{keV}$ by \citet{Galarraga_Espinosa_2021} for simulations and $\approx \SI{0.7}{keV}$ by \citet{Zhang_2024} for stacking observations), but is expected for a filament in the outskirts of a massive merging cluster, where the gas is already hotter, denser, and more dynamically disturbed, for example, through gravitational heating as material flows toward the cluster and through local compression between the main cluster and the infalling group (e.g., \citealt{Vurm_2023}). The filament temperature is lower by a factor of three than the temperature of A3266 between $R_{500}$ and $R_{200}$ toward the filament in the northwest ($\num{4.2}_{-1.7}^{+3.3}\,\mathrm{keV}$) and in a direction without a filament or group in the northeast ($\num{3.8}_{-1.4}^{+3.7}\,\mathrm{keV}$).

For the metallicity of the filament gas, we measured $Z = \num{0.07}_{-0.05}^{+0.09}\,{Z_\odot}$. Observational constraints on the metallicity of cosmic-web filaments remain scarce. Recent measurements by \citet{Veronica_2024} and \citet{Dietl_2024} found metallicities of $Z \approx \SI{0.1}{Z_\odot}$, consistent with our value. From the simulation side, the WHIM in circumcluster environments beyond $R_{200}$ in IllustrisTNG exhibits metallicities in the range $\SIrange{0.06}{0.13}{Z_\odot}$ \citep{Gouin_2023}, also consistent with our measurement.

\citet{Ilc_2024} reported that the simulated baryon physics strongly affects the metallicity of gas in filaments. They used the Dianoga set of simulations to measure WHIM metallicities of $Z = \SI{0.08(7)}{Z_\odot}$\footnote{The quoted uncertainties for the simulations reflect population scatter rather than statistical measurement errors.} with AGN feedback and $Z < \SI{0.03}{Z_\odot}$ without AGN feedback. In these simulations, AGN feedback enriches the WHIM primarily by driving metal-rich outflows from galaxies and cluster cores into the surrounding filamentary gas. Since the temperature of the filament gas in our system exceeds their definition of the WHIM, we also compared our result to their hot gas phase ($T > \SI{0.86}{keV}$), for which they reported metallicities of $\num{0.26}_{-0.14}^{+1.7}\,\textnormal{Z}_\odot$ with AGN feedback and $Z < \SI{0.04}{Z_\odot}$ without AGN feedback. 

Our metallicity measurement lies toward the lower end of the expected range including AGN feedback. The measurement is therefore compatible with scenarios in which metal enrichment of the filament gas is relatively inefficient or spatially limited. However, the current uncertainties do not allow us to distinguish robustly between different feedback models. More sensitive measurements of metallicities in filaments surrounding galaxy clusters are therefore crucial for constraining the efficiency of metal transport into filamentary gas and the baryonic processes that enable it in cosmological simulations.

Conversion of the normalization from the spectral fitting into physical quantities yields an electron number density of $n_{\rm e} = \SI{8.3(8)e-5}{\per\cubic\centi\meter}$ and a baryon overdensity of $\delta_{\rm b} = \num{320(30)}$. As mentioned in \cref{subsec:filament_gas}, the uncertainties here are only statistical, and the calculation assumed the filament to lie in the plane of the sky (inclination angle $i = \SI{0}{\degree}$). A redshift difference between A3266 and the NW group, that is, a finite inclination angle, would lead to a larger filament volume and therefore a lower gas density. The redshifts $z_{\rm A3266} = \num{0.0596(2)}$ \citep{Dehghan_2017} and $z_{\rm NW group} = 0.06_{-0.02}^{+0.01}$ (determined via our X-ray fit) agree with each other, but mainly because the uncertainty of the latter one is large. The redshift assigned to the NW group in the eROSITA cluster catalog (\cref{tab:group_props}) has no smaller uncertainties. To demonstrate the impact of the inclination angle, we computed it for an extreme scenario: taking the difference of the redshifts resulted in the upper limit $i < \SI{80}{\degree}$ and a 3D separation of cluster center to group center of $< \SI{24}{Mpc}$. This is much larger than the projected distance of $\SI{4}{Mpc}$. The resulting lower limits are $n_{\rm e} > \SI{3.1e-5}{\per\cubic\centi\meter}$ and $\delta_{\rm b} > \num{119}$. However, this extreme scenario is unlikely: First, the fact that we observe the density jump at the cold front suggests a small inclination angle. If the infall direction of the group were close to the line of sight, the cold front could not be observed. Second, \citet{Gatuzz_2024} measured no significant line-of-sight velocity component at the NW structure of A3266. Although their studied region is at $R_{500}$ of A3266 (see~Fig.~12 in \citealt{Gatuzz_2024}) and does not trace the filament or the NW group, it shows that there is no substantial gas inflow along the line of sight, making it plausible that the gas inflow occurs close to the plane of the sky. In order to report a reasonable systematic uncertainty, we chose a fiducial inclination angle of $i = \SI{20}{\degree}$ and varied it between $\SI{0}{\degree}$ and $\SI{50}{\degree}$. This resulted in a 3D separation (i.e., the height of the cylinder) of $L_{\rm center\ to\ center} = {4.5}_{-0.2}^{+2}\,\mathrm{Mpc}$ ($L_{R_{200}\text{ to }R_{200}} = {1.1}_{-0.1}^{+0.5}\,\mathrm{Mpc}$), $n_{\rm e} = {8}_{-2}^{+1}\times10^{-5}\,\mathrm{cm}^{-3}$, and $\delta_{\rm b} = {310}_{-80}^{+40}$. This is at the high end of the expectations for filaments (e.g., \citealt{Shull_2012} reports $\delta_{\rm b} < 316$ for the hot phase of the WHIM and \citealt{Martizzi_2019} defines $\delta_{\rm b} < 450$ (converted into our redshift) for the border between WHIM and halo gas. Therefore, our measurement is within expectations for a short dense filament within the cluster outskirts.

\section{Conclusions}
\label{sec:conclusions}
Our analysis revealed a complex network of galaxy groups and filaments in the outskirts of A3266, highlighting the dynamical interplay between large-scale structure and cluster assembly.

\begin{itemize}[label={-}]
\item We identified multiple galaxy groups in the near and far outskirts of A3266, several of which are connected by galaxy and/or gas filaments at different confidence levels. In particular, the NW group is definitively connected to the cluster, the NNW group is very likely connected, groups G and H are possibly connected (also to each other), and group F shows a potential connection in the galaxy distribution, but no detectable X-ray connection.
\item We detected a filament extending to the NW group at the $3.6\,\sigma$ level in excess of the combined cluster and group emission, with its axis aligned with the previously known NW structure, the NW group, group D, the cluster A3255, and the HRSC.
\item The NW group was identified as a cool-core group embedded within the filament and moving through the filament gas toward the main cluster. A cold front was observed at its leading edge, evidenced by a sharp surface-brightness discontinuity.
\item The filament gas between the main cluster and the NW group is hotter ($T = \num{1.2}_{-0.2}^{+0.3}\,$keV) and in the upper region of densities ($n_{\rm e} = {8}_{-2}^{+1}\times10^{-5}\,\mathrm{cm}^{-3}$, $\delta_{\rm b} = {310}_{-80}^{+40}$) than expected for pristine WHIM. This is consistent with its location in the cluster outskirts ($> R_{200}$ but $<3R_{200}$), where the gas has likely been disturbed by the complex merger history of A3266 and was further heated and compressed by the ongoing pre-merger of the NW group. The metallicity ($Z = \num{0.07}_{-0.05}^{+0.09}\,Z_\odot$) is consistent with recent observational and simulation-based estimates for filament gas in cluster environments and suggests relatively low levels of metal enrichment. This measurement adds to the still limited set of metallicity constraints for cosmic-web filaments. Because of the large uncertainties, no robust conclusions can be drawn regarding the role of different AGN feedback mechanisms, however.
\end{itemize}

\begin{acknowledgements}
JD received financial support for this research by the Federal Ministry of Education and Research (BMBF) and the Ministry of Culture and Science of the State of North Rhine-Westphalia (MWK) as part of TRA Matter and the Excellence Strategy of the federal and state governments. JD also acknowledges the International Max Planck Research School for Astronomy and Astrophysics (IMPRS A\&A) at the Universities of Bonn and Cologne for supporting him through a research contract. JD is a member of the IMPRS A\&A and the Bonn Cologne Graduate School (BCGS).
\\
JD acknowledges funding by the Deutsche Forschungsgemeinschaft (DFG, German Research Foundation) under Germany’s Excellence Strategy EXC 3037 - 533607693 - Our Dynamic Universe.
\\
YZ acknowledges support from the Chinese Scholarship Council (CSC) and the German Academic Exchange Service (DAAD).
\\
MCHY acknowledges the support from the Deutsche Forschungsgemeinschaft (DFG) through the grant FR 1691/2-1.
\\
This work is based on data from eROSITA, the soft X-ray instrument aboard SRG, a joint Russian-German science mission supported by the Russian Space Agency (Roskosmos), in the interests of the Russian Academy of Sciences represented by its Space Research Institute (IKI), and the Deutsches Zentrum für Luft- und Raumfahrt (DLR). The SRG spacecraft was built by Lavochkin Association (NPOL) and its subcontractors, and is operated by NPOL with support from the Max Planck Institute for Extraterrestrial Physics (MPE).

The development and construction of the eROSITA X-ray instrument was led by MPE, with contributions from the Dr. Karl Remeis Observatory Bamberg \& ECAP (FAU Erlangen-Nuernberg), the University of Hamburg Observatory, the Leibniz Institute for Astrophysics Potsdam (AIP), and the Institute for Astronomy and Astrophysics of the University of Tübingen, with the support of DLR and the Max Planck Society. The Argelander Institute for Astronomy of the University of Bonn and the Ludwig Maximilians Universität Munich also participated in the science preparation for eROSITA.

The eROSITA data shown here were processed using the eSASS software system developed by the German eROSITA consortium.
\\
This research has made use of the NASA/IPAC Extragalactic Database (NED) which is operated by the Jet Propulsion Laboratory, California Institute of Technology, under contract with the National Aeronautics and Space Administration.
\\
This work made use of the Python packages NumPy, SciPy, uncertainties, and Astropy. GitHub Copilot has been used for AI code completion and generation of short Python scripts.
\end{acknowledgements}

\bibliographystyle{aa}
\bibliography{list_bib}
\begin{appendix}
\section{Properties of galaxy clusters and groups in the field of view}
\label{sec:app_group_props}
\begin{table}[htbp]
\centering
\caption{Names and redshifts of groups and clusters in \cref{Fig:large-scale_overview}.}
\label{tab:group_props}

\begin{tabular}{llS}
\toprule
ID & Name & {$z$} \\
\midrule
A & A3266                     & \num{0.0591(4)}\textsuperscript{a} \\
B & 1eRASS J042724.0--600155  & \num{0.060(1)}\textsuperscript{a} \\
C & A3255                     & \num{0.0585(8)}\textsuperscript{a} \\
D & 1eRASS J041809.6--601105  & \num{0.0638(3)}\textsuperscript{b} \\
E & 1eRASS J042450.5--604655  & \num{0.07(1)}\textsuperscript{c} \\
F & 1eRASS J043838.1--615107  & \num{0.07(1)}\textsuperscript{c} \\
G & 1eRASS J041907.1--622537  & \num{0.07(1)}\textsuperscript{c} \\
H & 1eRASS J041412.7--623013  & \num{0.06(1)}\textsuperscript{c} \\
I & 1eRASS J040828.2--604926  & \num{0.053(2)}\textsuperscript{a} \\
J & RXGCC 171  & \num{0.0185(5)}\textsuperscript{d} \\
K & 1eRASS J041129.8--624537  & \num{0.0919(4)}\textsuperscript{b} \\
L & 1eRASS J042953.1--602619  & \num{0.31(2)}\textsuperscript{c} \\
M & 1eRASS J042901.1--601819  & \num{0.211(5)}\textsuperscript{c} \\
N & 1eRASS J042514.5--592025  & \num{0.124(6)}\textsuperscript{c} \\
O & 1eRASS J043219.3--594243  & \num{0.138(5)}\textsuperscript{c} \\
P & 1eRASS J043651.3--593140  & \num{0.09(1)}\textsuperscript{c} \\
\bottomrule
\end{tabular}
\tablefoot{The redshifts are taken from \citet{Kluge_2024}. Superscripts indicate the redshift origin:
\textsuperscript{a} spectroscopic from multiple member galaxies (\texttt{spec\_z\_boot}),
\textsuperscript{b} spectroscopic from central galaxy (\texttt{cg\_spec\_z}),
\textsuperscript{c} photometric (\texttt{photo\_z}),
\textsuperscript{d} literature (\texttt{lit\_z}).}
\end{table}

\newpage
\section{Additional surface brightness profiles}
\label{sec:app_SB_profs}
\begin{figure}[htbp]
    \centering
    \includegraphics[width=\hsize]{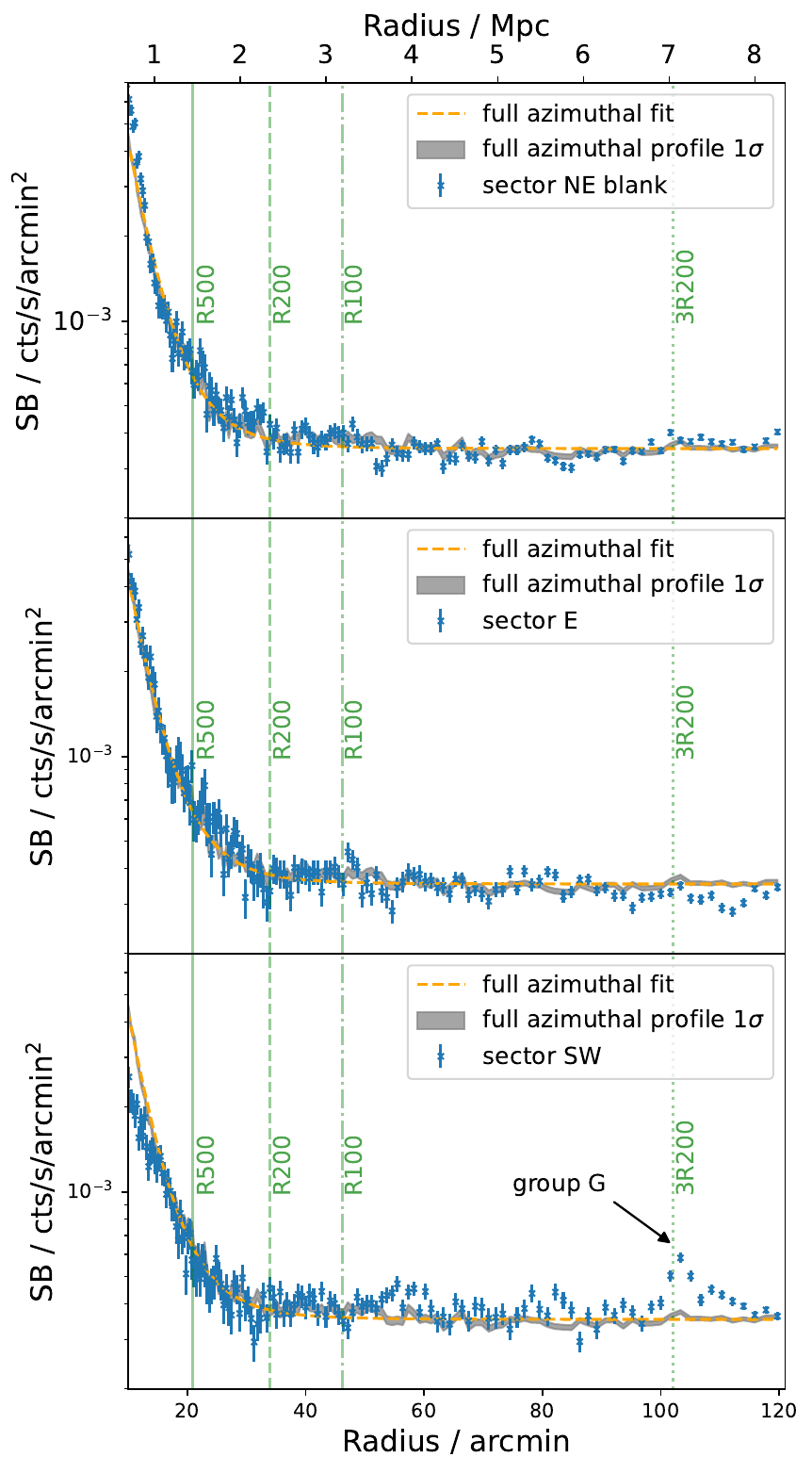}
    \caption{Additional sector profiles of A3266 following the definitions in \cref{Fig:wavelet_SB_defs}, with surrounding groups not excised. The azimuthally averaged fit and profile (surrounding groups excluded) are shown for comparison. The \texttt{NE blank} sector closely follows the average profile, similar to the \texttt{E} sector, which shows slightly enhanced emission up to $R_{100}$. The \texttt{SW} sector lies above the average profile. However, despite excision of the foreground cluster J, residual contamination cannot be ruled out. An excess associated with group G is visible at $\approx 3R_{200}$.
    }
    \label{Fig:SB3}
\end{figure}

\FloatBarrier
\newpage
\section{Spectral fitting results}
\label{sec:app_spectral_results}

\begin{table}[htbp]
\centering
\renewcommand{\arraystretch}{1.3}
\caption{Spectral results of the NW group and the filament region shown in \cref{Fig:spectra}.}
\label{tab:spectral_results}
\begin{tabular}{lccc}
\toprule
Region &
$T$ &
$Z$ &
Norm / Area \\
 &
in keV &
in $Z_\odot$ &
in $\mathrm{cm^{-5}\,arcmin^{-2}}$ \\
\midrule
$0$--$0.2\,R_{500}$ &
$0.82\pm0.04$ &
$0.17^{+0.03}_{-0.04}$ &
$(2.3\pm0.3)\times10^{-5}$ \\

$0.2$--$0.5\,R_{500}$ &
$0.90^{+0.05}_{-0.04}$ &
$0.18^{+0.06}_{-0.04}$ &
$(6.0\pm0.9)\times10^{-6}$ \\

$0.5\,R_{500}$--$R_{500}$ &
$1.02^{+0.09}_{-0.07}$ &
$0.17\pm0.08$ &
$1.9^{+0.6}_{-0.4}\times10^{-6}$ \\

$R_{500}$--$R_{200}$ &
$1.03^{+0.16}_{-0.11}$ &
$0.09\pm0.04$ &
$9^{+1}_{-2}\times10^{-7}$ \\

Filament &
$1.22^{+0.32}_{-0.15}$ &
$0.07^{+0.09}_{-0.05}$ &
$(1.2\pm0.3)\times10^{-6}$ \\
\bottomrule
\end{tabular}
\end{table}

\begin{table}[htbp]
\centering
\renewcommand{\arraystretch}{1.3}
\caption{Best-fit background normalizations for the regions shown in \cref{Fig:spectra}.}
\label{tab:background_results}
\begin{tabular}{lccc}
\toprule
Region &
\multicolumn{2}{c}{Norm$_{\mathrm{LHB / MWH}}$ / Area} &
Norm$_{\mathrm{AGN}}$ / Area \\
 &
 \multicolumn{2}{c}{ in $10^{-6}\,\mathrm{cm^{-5}\,arcmin^{-2}}$} &
 $\dagger$ \\
\midrule
$0$--$0.2\,R_{500}$ &
$2.6\pm0.1$ &
$0.71\pm0.03$ &
$4.5\pm0.3$ \\

$0.2$--$0.5\,R_{500}$ &
$2.6\pm0.1$ &
$0.71\pm0.03$ &
$4.5^{+0.3}_{-0.2}$ \\

$0.5\,R_{500}$--$R_{500}$ &
$2.6\pm0.1$ &
$0.72\pm0.03$ &
$4.6\pm0.2$ \\

$R_{500}$--$R_{200}$ &
$2.3\pm0.1$ &
$0.69^{+0.01}_{-0.03}$ &
$4.3^{+0.1}_{-0.2}$ \\

Filament &
$2.9\pm0.1$ &
$0.72^{+0.03}_{-0.02}$ &
$3.9^{+0.3}_{-0.2}$ \\
\bottomrule
\end{tabular}
{\\ \raggedright $\dagger$ in $10^{-7}$\,photons\,keV$^{-1}$\,s$^{-1}$\,cm$^{2}$\,arcmin$^{-2}$ at 1\,keV\par}
\end{table}

\begin{figure}[htbp]
    \centering
    \includegraphics[width=\linewidth]{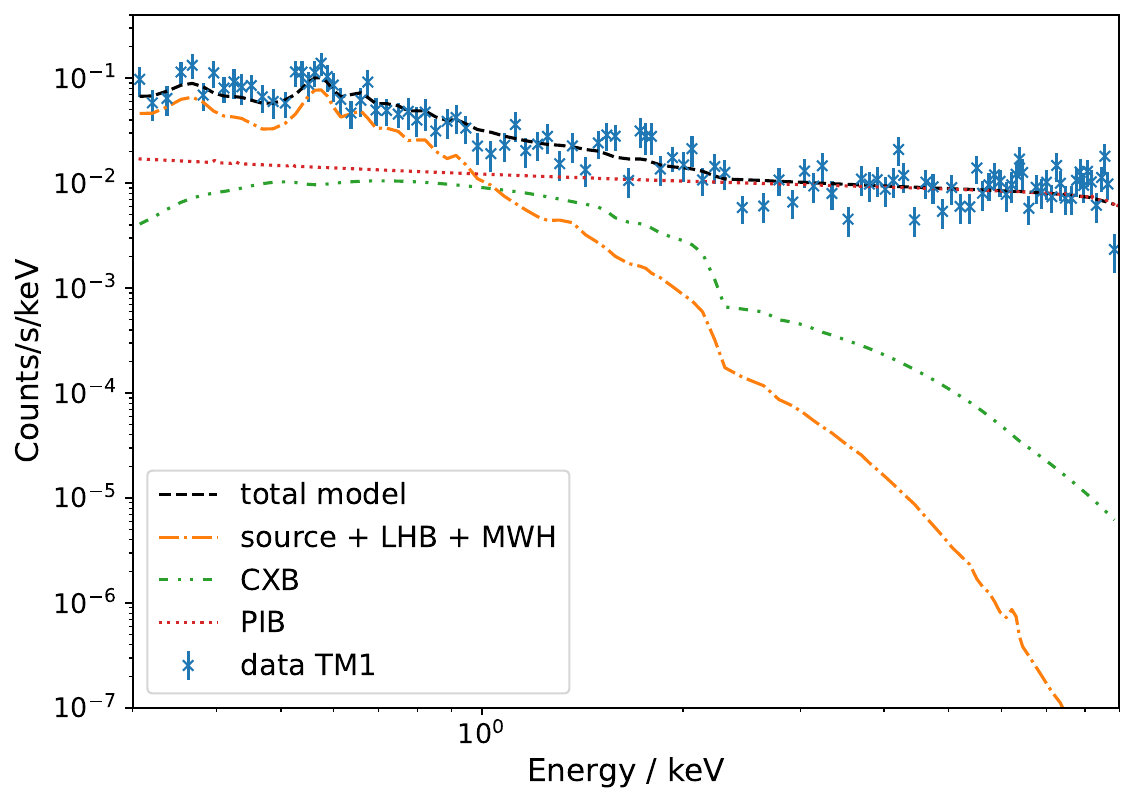}
    \caption{Example spectrum and best fit for TM1 for the filament region (cf.~\cref{Fig:spectral_regions}). The data is binned with 3 counts per bin for visualization.}
    \label{fig:TM1_spectrum}
\end{figure}

\end{appendix}

\end{document}